\documentclass[%
  reprint,
  superscriptaddress,
  amsmath,amssymb,
  aps,
  pra,
]{revtex4-2}

\usepackage{graphicx}
\usepackage{dcolumn}
\usepackage{bm}
\usepackage{bbold}
\usepackage[export]{adjustbox}
\usepackage{mathtools}
\usepackage[normalem]{ulem}
\usepackage{float}
\DeclarePairedDelimiter\bra{\langle}{\rvert}
\DeclarePairedDelimiter\ket{\lvert}{\rangle}
\DeclarePairedDelimiterX\braket[2]{\langle}{\rangle}{#1 \delimsize\vert #2}

\usepackage[utf8]{inputenc}
\usepackage[T1]{fontenc}
\usepackage{mathrsfs}
\usepackage{dsfont}
\usepackage{amssymb}
\usepackage{amsmath}
\usepackage{amsfonts}
\usepackage{amsthm}
\usepackage{xcolor}
\usepackage{hyperref}
\usepackage{comment}

\def\tr{\mathrm{tr}}

\newcommand{\ketvec}[1]{\ket{\vec{#1}}}
\newcommand{\bravec}[1]{\bra{\vec{#1}}}

\def\>{\rangle}
\def\<{\langle}

\begin{document}

\title{Equilibration and the eigenstate thermalization hypothesis as limits to observing macroscopic quantum superpositions}
\author{Gabriel Dias Carvalho}
\affiliation{Física de Materiais, Escola Politécnica de Pernambuco, Universidade de Pernambuco, 50720-001, Recife, PE, Brazil}
\email{gdc@poli.br}
\author{Pedro S. Correia}
\affiliation{Departamento de Ciências Exatas, Universidade Estadual de Santa Cruz, Ilhéus, Bahia 45662-900, Brazil}
\author{Thiago R. de Oliveira}
\affiliation{Instituto de Física, Universidade Federal Fluminense, Av. Litoranea s/n, Gragoatá 24210-346, Niterói, RJ, Brazil}
\date{\today}

\begin{abstract}

Macroscopic quantum superpositions are widely believed to be unobservable because large systems cannot be perfectly isolated from their environments. Here, we show that even under perfect isolation, intrinsic unitary dynamics in generic many-body systems, as analyzed through the eigenstate thermalization hypothesis and random matrix theory, can suppress the observable signatures of macroscopic coherence. Using the Greenberger-Horne-Zeilinger (GHZ) state as a representative example, we demonstrate that while fully correlated measurements can initially distinguish a macroscopic superposition from its corresponding classical mixture, generic many-body evolution renders them operationally indistinguishable for most times. By analyzing both distinguishability measures and established quantifiers of macroscopic quantumness, we find that equilibration not only hides coherence from accessible observables but also suppresses the corresponding signatures of macroscopic quantumness, in particular within the additive-local framework considered here. These results identify unitary thermalization, independent of environmental decoherence, as a fundamental mechanism that limits the observation of macroscopic quantum effects.

\end{abstract}

\maketitle

\section{Introduction and Motivation}

Since the early days of quantum mechanics, there has been no consensus regarding the measurement problem, particularly the dynamical mechanism underlying the apparent collapse of quantum superpositions into definite outcomes \cite{maudlin1995three}. 
The difficulty lies in reconciling the reversible Schrödinger evolution with the seemingly irreversible emergence of classicality through measurement. Measurements thus mark the point at which quantum features vanish from everyday experience.

A widely accepted explanation for this disappearance is decoherence \cite{decoherence_RMP}, whereby entanglement with an environment suppresses interference terms in the system. As environmental coupling drives this process, perfectly isolated systems, in principle, retain coherence indefinitely. While experiments confirm long-lived coherence in small, well-controlled systems, macroscopic superpositions remain experimentally inaccessible.

Even assuming an ideal, perfectly isolated system, macroscopic quantum states face fundamental limitations. Distinguishing a macroscopic superposition from its classical mixture requires measurement resources that scale at least quadratically with system size due to quantum reference-frame constraints \cite{Skotiniotis2017macroscopic}. 
Likewise, preparing such states requires resources that also scale unfavorably with system size, and even infinitesimal local noise can destroy their coherence \cite{LopezIncera2019allmacroscopic}. These considerations imply that macroscopic quantum effects are constrained not only by practical difficulties but also by fundamental scaling laws. Another core problem is identifying an observable that can distinguish a macroscopic superposition from a mixture. It has been shown, for instance, that no local additive observable can achieve this \cite{shimizu2005}.

This fragility, however, is not solely a matter of state preparation or measurement capability--it is also inherently dynamical. Even if one could prepare a macroscopic superposition, perform such high-demand measurements, and achieve complete isolation, the system may still undergo nontrivial unitary evolution driven by its own Hamiltonian. As we show in this work, such intrinsic 
dynamics can render a macroscopic superposition (e.g., a GHZ state) and its corresponding classical mixture operationally indistinguishable for experimentally accessible observables. The underlying mechanism is the same as the equilibration and thermalization of isolated quantum systems, where even an isolated quantum system in a pure state, evolving unitarily, may appear as a thermal mixed state due to entanglement between its constituents. 
In fact, recent works have already proposed a connection between quantum thermalization and the emergence of classicality  \cite{demelo2025, schwarzhans2025}, a perspective we further explore here
\footnote{Previously, some authors have proposed that even without an external environment, inaccessible internal degrees of freedom can induce a form of self-induced decoherence  \cite{castagnino2013,castagnino2008,castagnino2000}. This mechanism has been compared with decoherence in \cite{Schlosshauer05}; its connection to the equilibration of isolated quantum systems remains unclear.}. 

More concretely, we show that even for non-local observables, purely unitary dynamics--through the process of equilibration--can suppress detectable differences between the GHZ state and its corresponding classical mixture. While highly non-local, fully correlated measurements may reveal coherence at short times, this detectability fades under generic many-body evolution. For generic, non-degenerate many-body Hamiltonians, the expectation values for the two states converge in the long-time regime for the class of observables considered in Sec. II.

To quantify the loss of coherence, we employ a well-established measure of macroscopic quantumness \cite{RevFrowis, shimizu2005, Morimae2005}. We examine how quantum fluctuations scale with $N$ for both magnetization-like and correlation-like observables. Our analysis confirms that the coherence measure from \cite{Morimae2005} diminishes upon equilibration. 
Through both analytical arguments and numerical simulations we show that equilibration suppresses off-diagonal elements in the energy basis. Consequently, observable differences between macroscopic superpositions and statistical mixtures are suppressed over time.
This establishes equilibration as key intrinsic mechanism for the operational disappearance of quantum coherence at large scales.

The remainder of this article is organized as follows: section \ref{sec:analytical} analytically evaluates distinguishability between macroscopic superpositions and mixtures via magnetization-like and correlation-like observables, at both short and long times. Section \ref{sec:measures} introduces coherence measures to quantify this suppression, and section \ref{sec:conclusions} concludes with implications and future directions.

\section{Indistinguishability and observables}
\label{sec:analytical}

Schrödinger's cat paradox remains the iconic illustration of the tension between the quantum and classical descriptions of nature. Schrödinger showed that applying quantum mechanics to a combined microscopic-macroscopic system can, in principle, lead to a superposition of macroscopic distinct outcomes-typically described as a cat being simultaneously dead and alive. The paradox arises from the fact that such macroscopic superpositions have never been observed in practice.

This naturally raises an operational question: how can we verify that a macroscopic system is in a coherent superposition rather than in a classical statistical mixture? In microscopic systems, interference experiments such as the double slit provide a clear answer, but extending such strategies to macroscopic systems is highly nontrivial. Mathematically, detecting coherence requires measuring an observable $A$ that is not diagonal on the basis in which the coherence is encoded.

To make this concrete, consider a paradigmatic representative of a macroscopic superposition; the GHZ state:
\begin{equation}
\ket{\psi_\mathrm{GHZ}} =\bigl(\ketvec{0} + \ketvec{1}\bigr)/\sqrt{2} ,
\end{equation}
with $\ketvec{0} \equiv \ket{0}^{\otimes N}$ and $\ketvec{1} \equiv \ket{1}^{\otimes N}$. The question is whether one can verify that we actually have such a state. Verifying a complete quantum state is a very complex task and is vital in many applications, such as quantum computing \cite{Flammia2011,Guhne2009}. However, let us simplify our problem and question if there is an observable that can distinguish $\varrho_{\mathrm{GHZ}}=|\psi_{\mathrm{GHZ}}\>\<\psi_{\mathrm{GHZ}}|$ from the corresponding classical state 
\begin{equation}
\rho_{\mathrm{mix}}= \bigl(\ketvec{0}\bravec{0} + \ketvec{1}\bravec{1}\bigr)/2.
\end{equation}
The operational distinguishability of these two states hinges on measuring an observable $A$ for which the expectation value differs:
\begin{align}
\Delta\langle A\rangle &= \operatorname{tr}\bigl[A(\rho_{\mathrm{GHZ}} - \rho_{\mathrm{mix}})\bigr]
= \operatorname{Re}\bigl(\bravec{0}A\ketvec{1}\bigr).
\label{eq:DifferenceKey}
\end{align}
Thus, any measurable difference is encoded entirely in the off-diagonal coherent element $\bravec{0}A\ketvec{1}$.

A natural first candidate is the local, additive observable corresponding to total magnetization
\begin{align}
A_{L}\;=\;\sum_{j=1}^{N}\sigma_{\hat n}^{(j)},
\label{eq:magoperator_main}
\end{align}
where $\sigma_{\hat n}^{(j)}=\mathbb 1^{\otimes(j-1)}\,\!\otimes\sigma_{\hat n}\! \,\otimes\mathbb 1^{\otimes(N-j)}$ denotes a Pauli operator pointing along some fixed direction $\hat n$ and acting solely on the $j$-th qubit. However, this distinction fails
\begin{align}
\Delta\!\left\langle A_{L}\right\rangle
&=\operatorname{Re}\,\!\Bigl(\bra{\vec 0}A_{L}\ket{\vec 1}\Bigr)
   =\operatorname{Re}\,\!\Bigl(\sum_{j=1}^{N}\bra{\vec 0}\sigma_{\hat n}^{(j)}\ket{\vec 1}\Bigr)=0,
\label{eq:DeltaA2}
\end{align}
since \( \bra{\vec 0}\sigma_{\hat n}^{(j)}\ket{\vec 1}=\bra{0}\sigma_{\hat n}\ket{1}\prod_{k\neq j}\langle 0 | 1 \rangle=0\). No single-spin operator can convert $\ket{\vec 0}$ into $\ket{\vec 1}$ without leaving orthogonal components, making the coherence invisible to such measurements.
Therefore, any local additive observable of the form~\eqref{eq:magoperator_main} fails to distinguish the GHZ state from its classical mixture.  

In contrast, a fully correlated, non-local observable such as
\begin{align}
A_{NL}\;=\;\sigma_{\hat n}^{\otimes N},
\label{eq:coroperator_main}
\end{align}
can distinguish the states. Its expectation value difference is  
\begin{align}
\Delta\!\langle A_{NL}\rangle
  &=\operatorname{Re}\!\bigl(\bra{\vec 0}\sigma_{\hat n}^{\otimes N}\ket{\vec 1}\bigr)
   =\operatorname{Re}\!\Bigl[\bigl(\bra 0\sigma_{\hat n}\ket 1\bigr)^{\!N}\Bigr].
\label{eq:DeltaA1}
\end{align}
For a generic unit vector $\hat n=(n_x,n_y,n_z)$, one finds  $\bra 0\sigma_{\hat n}\ket 1 = n_x+in_y$ and $\bigl|\bra 0\sigma_{\hat n}\ket 1\bigr|=\sqrt{1- n_z^{2}}\le 1$.
Hence, $n_z=\epsilon$ implies that $\Delta\!\langle A_{NL}\rangle \rightarrow 0$ exponentially with $N$. Thus, distinguishing the GHZ state from its mixture in the macroscopic limit requires not only a very non-local measurement but also extremely precise alignment of the measurement direction. Otherwise, the detectable signal becomes exponentially small, requiring an impractically large number of measurement repetitions to resolve.

\section{Equilibration}

Even if one could prepare the GHZ state and perform the required nonlocal, high-precision measurement, the situation remains problematic. In a realistic laboratory setting the system will evolve over time and may decohere due to imperfect isolation. More importantly, even a perfectly isolated system evolving strictly under unitary dynamics does not guarantee the persistence of a macroscopic superposition. The complex internal Hamiltonian of a many-body system acts continuously, and we generally lack the ability to control or reverse this intrinsic evolution. As a result, macroscopic superpositions are dynamically unstable: for most times, a given observable $A$ may fail to distinguish $\rho_{\mathrm{GHZ}}$ from $\rho_{\mathrm{mix}}$ leading to $\Delta\langle A(t)\rangle \ll 1$ for the overwhelming majority of times t. This behavior may originate from the equilibration and thermalization phenomena known for isolated quantum systems.

Let us briefly review key results on the equilibration of isolated quantum systems. Consider an initial state $|\psi\>=\sum_n c_n |E_n\>$ evolving unitarily under the Hamiltonian $H=\sum_n E_n |E_n\>\<E_n|$. Although the unitary evolution never ceases--the system never approaches a true stationary state--an observer with access only to a limited set of observables $\mathcal{O}$ may nonetheless find that the system appears equilibrated. Specifically, for any observable $\hat{O} \in \mathcal{O}$ the expectation value $\<\hat{O}(t)\>=\<\psi(t)|\hat{O}|\psi(t)\> = \text{Tr}[\rho(t)\hat{O}]$ remains close to its time average for most values of t. The time average is defined as
\begin{equation}
\overline{\<\hat{O}\>} =\lim_{T\to\infty}\frac1T\int_{0}^{T}\!dt\,\<\hat{O}(t)\>,
\end{equation}
which is time invariant. To quantify the equilibration, one considers the average size of the fluctuations:
\begin{equation}
\sigma_{\hat{O}} = \overline{(\<\hat{O}(t)\> - \overline{\<\hat{O}\>})^2},    
\end{equation}
with the overline representing the time average. If $\sigma_{\hat{O}} \ll 1$ then $\<\hat{O}(t)\> \approx \overline{\<\hat{O}\>}$ for almost all times $t$. Under very general conditions, the following bound holds \footnote{Recent results were obtained in \cite{Reimann08, Linden08} Later, it was
found that von Neumann already obtained similar results. A not-so-recent but also important and usually
forgotten reference is \cite{Tasaki98}}:
\begin{equation}
\sigma_{\hat{O}} \leq ||\hat{O}|| \text{Tr}[\overline{\rho}^2]
\end{equation}
where $||\hat{O}||$ is the operator norm of $\hat{O}$ and $\overline{\rho}=\sum_n |c_n|^2 |E_n\>\<E_n|$ the time average state. Note that $\overline{\rho}$ is the initial state dephased in the $H$ basis. Equilibration thus requires that $\overline{\rho}$ have low purity. In many-body systems, this condition is naturally satisfied: energy levels are exponentially dense, so experimentally realistic initial states typically populate many eigenstates \cite{Reimann08}. Consequently, one usually has $\text{Tr}[\overline{\rho}^2] \sim 1/d$, where $d$ is the Hilbert space dimension, which grows exponentially with particle number. Note that the system never truly converges to $\overline{\rho}$; rather, its behavior is such that $\<\hat{O}(t)\>$ is extremely likely to be close to the average value at almost all times; it is a probabilistic convergence. 

The result above is valid for Hamiltonians without "degenerate gaps": $ E_k - E_l = E_m - E_n $ only if $E_k = E_m; E_l = E_n$ or $E_k = E_l ; E_m = E_n$. This condition is believed to hold generically for interacting many-body systems and is well supported by numerical evidence. The results also extend to Hamiltonians with a non-exponentially large number of degenerate gaps or finite-time windows \cite{Short2012}, to infinite-dimensional systems \cite{Reimann2012}, and even to certain classes of time-dependent Hamiltonians \cite{Passos25}.

Returning to our problem, even if an observable $A$ can in principle distinguish $\rho_{\mathrm{GHZ}}$ from $\rho_{\mathrm{mix}}$ by yielding a finite $\Delta \<A\>$ at some moment, the intrinsic unitary evolution of an isolated many-body system may render $\Delta \<A(t)\>$  extremely small for almost all t. To show this, two ingredients are needed: i) equilibration of the signal: $\Delta \<A(t)\>$ must exhibit small temporal fluctuations, meaning its time variance is small. ii) Small long-time signal: the time average value $\overline{\Delta \<A\>}$ must itself be very small.
When both conditions are satisfied, the GHZ state and its corresponding mixture become operationally indistinguishable for practically all times during the unitary evolution.

\section{Analytical Results under ETH Hypotheses}

We will make the calculations in the Heisenberg picture; thus, the observable $A$ evolves in time, while $\rho_{\mathrm{mix}}$ and $\rho_{\mathrm{GHZ}}$ remain stationary. Hence, \(\Delta\<A(t)\> = \operatorname{Re}\!\bigl(\bra{\vec 0}A(t)\ket{\vec 1}\bigr),\\
\)

\begin{align}
\bra{\vec 0}A(t)\ket{\vec 1}
  = \sum_{m,n} e^{-i(E_n-E_m)t}\,
    \bra{E_m}A\ket{E_n}\,
    \langle\vec 0|E_m\rangle\,
    \langle E_n|\vec 1\rangle .
\end{align}

Defining
\(\Tilde{A}_{m,n} =
\bra{E_m}A\ket{E_n}
\langle\vec 0|E_m\rangle
\langle E_n|\vec 1\rangle\),
we can write
\(\Delta\<A(t)\> = \operatorname{Re}\!\left[
\sum_{m,n} e^{-i(E_n-E_m)t}\Tilde{A}_{m,n}
\right]\).
Letting
\(f(t) = \sum_{m,n} e^{-i(E_n-E_m)t}\Tilde{A}_{m,n}\),
it follows that \(\Delta\<A(t)\> = \tfrac{1}{2}(f(t) + f^*(t))\).

The long-time average of the squared fluctuation is
\begin{equation}
\overline{(\Delta\<A(t)\>)^2}
= \tfrac{1}{2}\,\overline{|f(t)|^2}
+ \tfrac{1}{2}\,\operatorname{Re}\!\bigl[\overline{f(t)^2}\bigr].
\end{equation}
Expanding \(f(t) = \sum_{m,n} e^{-i(E_n-E_m)t}\Tilde{A}_{m,n}\), 
\(f^*(t) = \sum_{p,q} e^{+i(E_q-E_p)t}\Tilde{A}_{p,q}^*\nonumber\nonumber\), gives
\begin{equation}
|f(t)|^2
= \sum_{m,n,p,q}
e^{-i[(E_n-E_m)-(E_q-E_p)]t}\,
\Tilde{A}_{m,n}\Tilde{A}_{p,q}^*.
\end{equation}
In the time average, only stationary terms survive. For non-degenerate energy gaps,
\(E_n - E_m = E_q - E_p\) implies \(n=q\) and \(m=p\),
hence \(\overline{|f(t)|^2} = \sum_{m,n}|\Tilde{A}_{m,n}|^2.\) Similarly, for
\(f(t)^2 =
\sum_{m,n,p,q}
e^{-i[(E_n-E_m)+(E_q-E_p)]t}\Tilde{A}_{m,n}\Tilde{A}_{p,q}\),
the stationary condition
\((E_n-E_m)+(E_q-E_p)=0\)
requires exchanged indices \((p,q)=(n,m)\),
leading to
\(\overline{f(t)^2} = \sum_{m,n}\Tilde{A}_{m,n}\Tilde{A}_{n,m}\).
Taking the real part and substituting both results yields
\begin{align}
\overline{(\Delta\<A(t)\>)^2}
= \tfrac{1}{2}\sum_{m,n}|\Tilde{A}_{m,n}|^2
+ \tfrac{1}{2}\sum_{m,n}\operatorname{Re}
   \bigl(\Tilde{A}_{m,n}\Tilde{A}_{n,m}\bigr).
    \label{eq:TimeAvg}
\end{align}

Using the eigenstate thermalization hypothesis (ETH) together with assumptions from random matrix theory, we will show that both
$\overline{\Delta \<A(t)\>}$ and $\overline{(\Delta \<A(t)\>)^2}$ go to zero for large systems. The arguments are very similar, and therefore we only present them for $\overline{(\Delta \<A(t)\>)^2}$. Here and throughout, $O(\cdot)$ denotes standard Big-O, or order-of, notation.

For generic, non-integrable many-body Hamiltonians, random matrix theory suggests that energy eigenstates $|E_m\rangle$ exhibit effectively random phases in the computational basis. As a consequence, typical overlaps scale as $|\langle \vec{0}|E_m\rangle|,\, |\langle E_m|\vec{1}\rangle| \sim O(2^{-N/2})$ \cite{DAlessio2016}. In addition, for physical observables, to which statistical mechanics applies, ETH implies that the typical off-diagonal matrix elements scale as \(|\langle E_m|A|E_n\rangle| \sim O(e^{-S/2}) \, (m\neq n), \) where $S$ is the entropy associated with the relevant energy window, together with $|\langle E_m|A|E_m\rangle|$ proportional to the expectation value of the corresponding microcanonical ensemble \cite{DAlessio2016}. For the generic nonintegrable Hamiltonians considered here, the GHZ state and its corresponding mixture have support over an exponentially large fraction of the spectrum, and as a consequence, their unitary time evolution leads to equilibration, with an effective energy distribution centered near the middle of the spectrum. In this regime, the entropy scales as $S \approx N \ln 2$, so that $e^{-S/2} = 2^{-N/2}$. Combining these estimates gives the typical size of the coefficients $\Tilde{A}_{mn}$ in Eq.~\eqref{eq:TimeAvg}:
\begin{align}
    |\Tilde{A}_{mn}|^2 \sim O(2^{-3N}).    
\end{align}

Since the sum in Eq.~\eqref{eq:TimeAvg} contains $O(2^{2N})$ terms, its contribution scales as $\overline{(\Delta\langle A(t)\rangle)^2}   \sim O(2^{-N})$, which shows that equilibration suppresses the distinguishability signal for the observables considered here.

For the additive operator $A_L=\sum_{j=1}^N \sigma_{\hat n}^{(j)}$ one has $\|A_L\|\le N$, yielding
\begin{align}
    \overline{(\Delta\langle A_L(t)\rangle)^2}
    \sim O(2^{-N})
    \xrightarrow{N\to\infty} 0.    
\label{eq:averageDeltaLocal}
\end{align}

The estimate leading to Eq. (\ref{eq:averageDeltaLocal}) assumes that the relevant energy window has extensive entropy. In the middle of the spectrum, \(S\simeq N\ln 2\), so that \(e^{-S/2}\simeq 2^{-N/2}\). Away from the middle of the spectrum, the same argument should instead be expressed in terms of the ETH factor \(e^{-S(E)/2}\), where \(S(E)\) is the microcanonical entropy of the relevant energy window. Thus, whenever the relevant energy window has extensive entropy, \(S(E)=O(N)\), the distinguishability signal remains exponentially suppressed, although with an energy-dependent exponent. Near spectral edges, however, the effective dimension of the relevant energy window may be much smaller, and the ETH/RMT estimate used above is not guaranteed to apply \cite{DAlessio2016}.

For the fully correlated operator \(A_{NL}=\sigma_{\hat n}^{\otimes N}\), the norm is intensive, \(\|A_{NL}\|=1\). For spin-\(\tfrac{1}{2}\) constituents this follows from the fact that each \(\sigma_{\hat n}\) is a dimensionless Pauli operator with eigenvalues \(\pm 1\), and the same normalization can be extended to higher spins by using \(\sigma_{\hat n}^{(s)}=S_{\hat n}/(s\hbar)\). Thus, despite being highly non-local, \(A_{NL}\) remains a bounded observable. Its ETH scaling, however, depends on how it is represented in the computational configuration basis. The physically relevant case for distinguishing the GHZ state from its classical mixture without an exponentially small initial signal corresponds to equatorial directions, \(n_z=0\). In this case, \(A_{NL}\) maps each configuration to a unique complementary configuration and is therefore a monomial operator, with exactly one nonzero entry in each row and column.

This structure places the equatorial \(A_{NL}\) directly within the Behemoth-operator framework of Ref.~\cite{Behemoth}. A Behemoth is an elementary configuration-basis operator of the form \(|\alpha\rangle\langle\beta|\), containing a single nonzero matrix element. Hence, an operator with \(M\) nonzero entries in the configuration basis can be viewed as a sum of \(M\) Behemoths. Ref.~\cite{Behemoth} shows that the width of the corresponding eigenstate matrix-element distribution scales as \(\sqrt{M}D^{-1}\), where \(D=2^N\) is the Hilbert-space dimension. A single Behemoth therefore exhibits the super-ETH scaling \(D^{-1}\), whereas an operator with \(M=O(D)\) nonzero entries exhibits the usual ETH scaling \(D^{-1/2}\). Since the equatorial \(A_{NL}\) has exactly \(D\) nonzero entries, it belongs to this ETH-scaling class.

For a generic direction with $n_z\neq 0$, however, $A_{NL}$ becomes a weighted sum of Behemoths connecting configurations at all Hamming distances; in this case, the initial GHZ--mixture contrast is already exponentially suppressed, since $|\langle 0|\sigma_{\hat n}|1\rangle|^N = (1-n_z^2)^{N/2}$, rendering these directions operationally irrelevant for macroscopic distinguishability. In all cases, $||A_{NL}||=1$, and for the equatorial directions \cite{Behemoth} justifies the ETH-consistent scaling adopted here. Hence,
\begin{align}
    \overline{(\Delta\langle A_{NL}(t)\rangle)^2}
    \sim O(2^{-N})
    \xrightarrow{N\to\infty} 0.    
\end{align}

Thus, both local and fully correlated observables lose their distinguishing power after equilibration, within the class of observables and initial conditions considered here.

\section{Numerics}

To illustrate the analytical results and the equilibration mechanism discussed above, we numerically study a chaotic spin chain governed by an extended Heisenberg–XYZ Hamiltonian. The model includes next-nearest neighbor interactions, transverse ($h_x$) and longitudinal ($h_z$) magnetic fields, and a local defect
of strength $e$ that explicitly breaks integrability. The Hamiltonian is
\begin{equation}
\begin{split}
H = &\sum_{i} \big[ J_{1} ( S_{i,x}S_{i+1,x} + S_{i,y}S_{i+1,y} + d\, S_{i,z}S_{i+1,z}) \\
&+ J_{2}\, (S_{i,x}S_{i+2,x} + S_{i,y}S_{i+2,y} + d\, S_{i,z}S_{i+2,z}) \\
&+ h_{x}\, S_{i,x} + h_{z}\, S_{i,z} + e\, S_{0,x} \big],  
\label{Hamiltonian}
\end{split}
\end{equation}
where $S_{i,\alpha}=\tfrac{1}{2}\sigma_{i,\alpha}$ ($\alpha\in\{x,y,z\}$) denotes the spin-$\tfrac{1}{2}$ operator acting on site $i$; we set $\hbar=1$. The parameters $J_{1}$ and $J_{2}$ are coupling strengths and $d$ an anisotropy parameter. This Hamiltonian is known to exhibit chaotic dynamics for generic parameter choices, ensuring that ETH-type behavior is expected.

\begin{figure}[h]
    \centering
    \begin{minipage}[t]{0.48\textwidth}
        \centering
        \includegraphics[width=\linewidth]{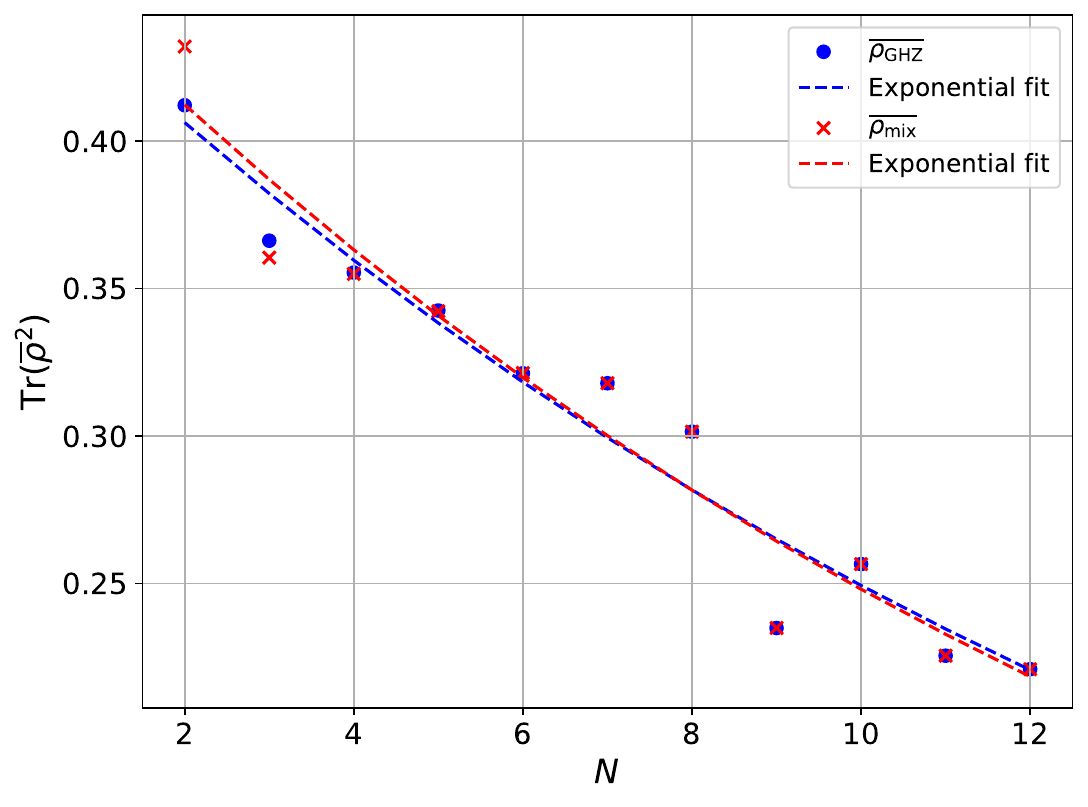}
    \end{minipage}
    \hfill
    \begin{minipage}[t]{0.48\textwidth}
        \centering
        \includegraphics[width=\linewidth]{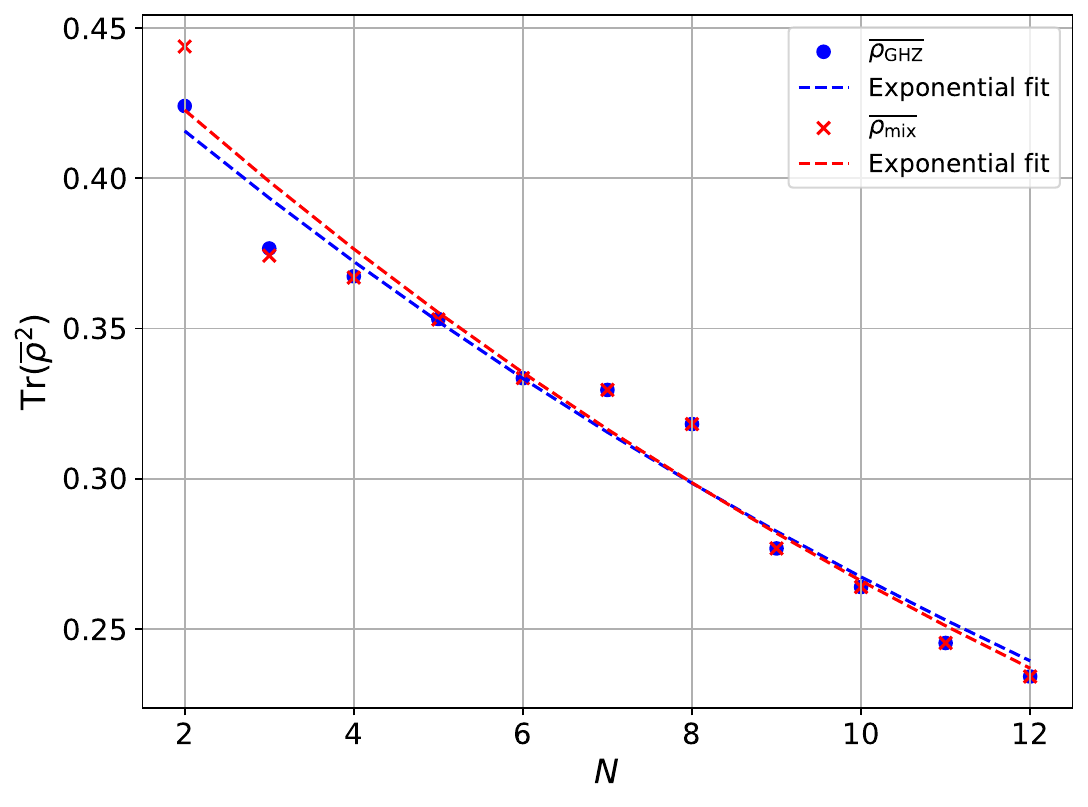}
    \end{minipage}
    \caption{Purity (inverse of the effective dimension) for $\overline{\rho_{\mathrm{GHZ}}}$ and $\overline{\rho_{\mathrm{mix}}}$ under different parameter values. On the top, $h_z$, $h_x$, $J_1$, $J_2$, $d$, and $e$ are fixed at 0.6, 0.2, 1.0, 1.35, 0.5, and 0.2, respectively. On the bottom, $h_z$, $h_x$, $J_1$, $J_2$, $d$, and $e$ are fixed at 0.6, 0.2, 1.0, 1.35, 0.5, and 0.1, respectively.}
    \label{fig:inv_eff_combined}
\end{figure}

To verify that $\Delta \<A(t)\>$ equilibrates, we analyze the equilibration of $A(t)$ for each initial state $\rho_{\mathrm{GHZ}}$ and $\rho_{\mathrm{mix}}$. We compute the purity of the time average state: $\text{Tr}[\overline{\rho_{\mathrm{GHZ}}}^2]$ and $\text{Tr}[\overline{\rho_{\mathrm{mix}}}^2]$. These purities are shown in Fig. \ref{fig:inv_eff_combined}. Their small values indicate that both initial states equilibrate under the dynamics generated by Hamiltonian (\ref{Hamiltonian}), validating the comparison of their time average observables. This behavior is robust: similar results were found in all $25$ parameter sets studied.

\begin{figure}[t]
    \includegraphics[width=\linewidth]{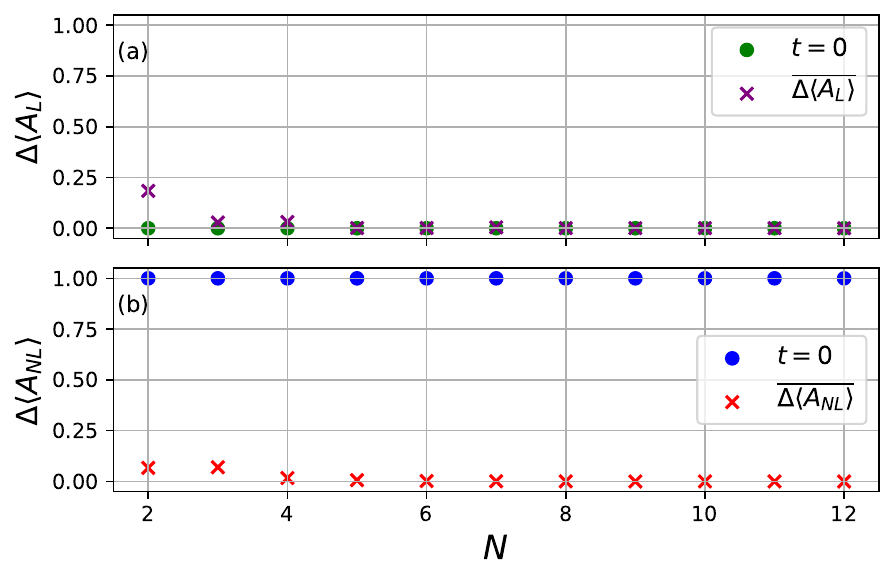} 
    \caption{Maximization over $\hat{n}$ of the differences $\tr\bigl[A (\rho_{\mathrm{GHZ}} - \rho_{\mathrm{mix}})\bigr]$, time equal $0$, and $\tr\bigl[A (\overline{\rho_{\mathrm{GHZ}}} - \overline{\rho_{\mathrm{mix}}})\bigr]$, time $t \rightarrow \infty$, for $A$ in \ref{eq:magoperator_main} (a) and \ref{eq:coroperator_main} (b). Parameters $h_{z}$, $h_{x}$, $J_{1}$, $J_{2}$, $d$, and $e$, are fixed with values 0.6, 0.2, 1.0, 1.35, 0.5, and 0.2, respectively. We found a similar behavior for all sets of parameters studied.}    \label{graph:diflineone}
\end{figure}

\begin{figure}[t]
    \includegraphics[width=\linewidth]{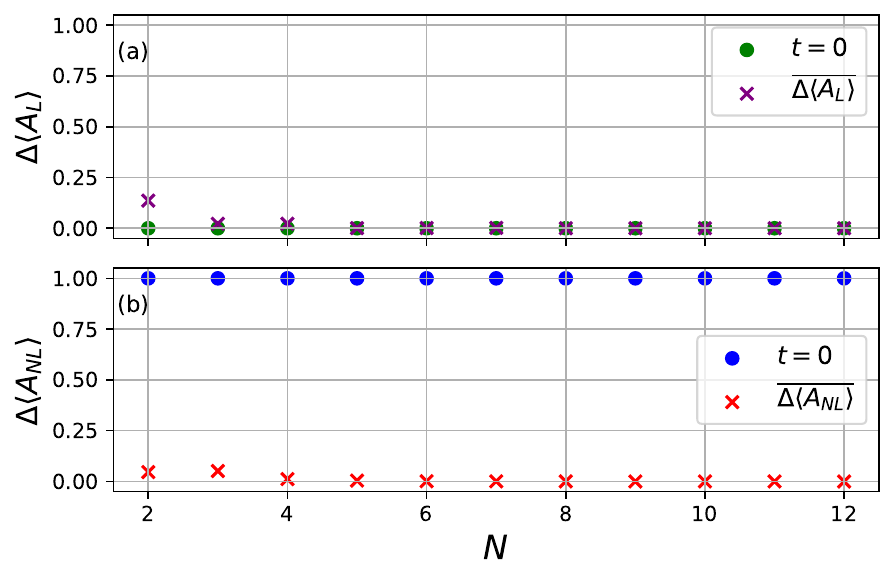} 
    \caption{Maximization over $\hat{n}$ of the differences $\tr\bigl[A (\rho_{\mathrm{GHZ}} - \rho_{\mathrm{mix}})\bigr]$, time equal $0$, and $\tr\bigl[A (\overline{\rho_{\mathrm{GHZ}}} - \overline{\rho_{\mathrm{mix}}})\bigr]$, time $t \rightarrow \infty$, for $A$ in \ref{eq:magoperator_main} (a) and \ref{eq:coroperator_main} (b). Parameters $h_{z}$, $h_{x}$, $J_{1}$, $J_{2}$, $d$, and $e$, are fixed with values 0.6, 0.2, 1.0, 1.35, 0.5, and 0.1, respectively. We found a similar behavior for all sets of parameters studied.}    \label{graph:diflastline}
\end{figure}

Next, we evaluate $\Delta \<A\>$ at $t=0$ and its time average $\overline{\Delta \<A\>}$ for both the local magnetization operator Eq. (\ref{eq:magoperator_main}) and the fully correlated operator Eq. (\ref{eq:coroperator_main}). For each $N$, we maximize over all possible measurement directions $\hat{n}$. We performed this analysis for $25$ different choices of Hamiltonian parameters; two representative cases are depicted in Figs. \ref{graph:diflineone} and \ref{graph:diflastline}. The numerical results corroborate the analytical predictions. As expected, the local operator $A_L$ fails to distinguish the GHZ state from the mixture-neither at $t=0$ nor at most later times. The non-local operator $A_{NL}$ can distinguish the two states at $t=0$, provided the measurement direction is properly aligned, but loses its discriminating power at later times for almost all $t$, consistent with equilibration under unitary dynamics.

This behavior holds across all parameter sets examined, reinforcing the conclusion that equilibration alone--without external decoherence--renders macroscopic superpositions operationally indistinguishable from their corresponding classical mixtures.

\section{Measures of Macroscopic Superpositions}

\label{sec:measures}

\begin{figure*}
    \centering
    \includegraphics[width=\textwidth]{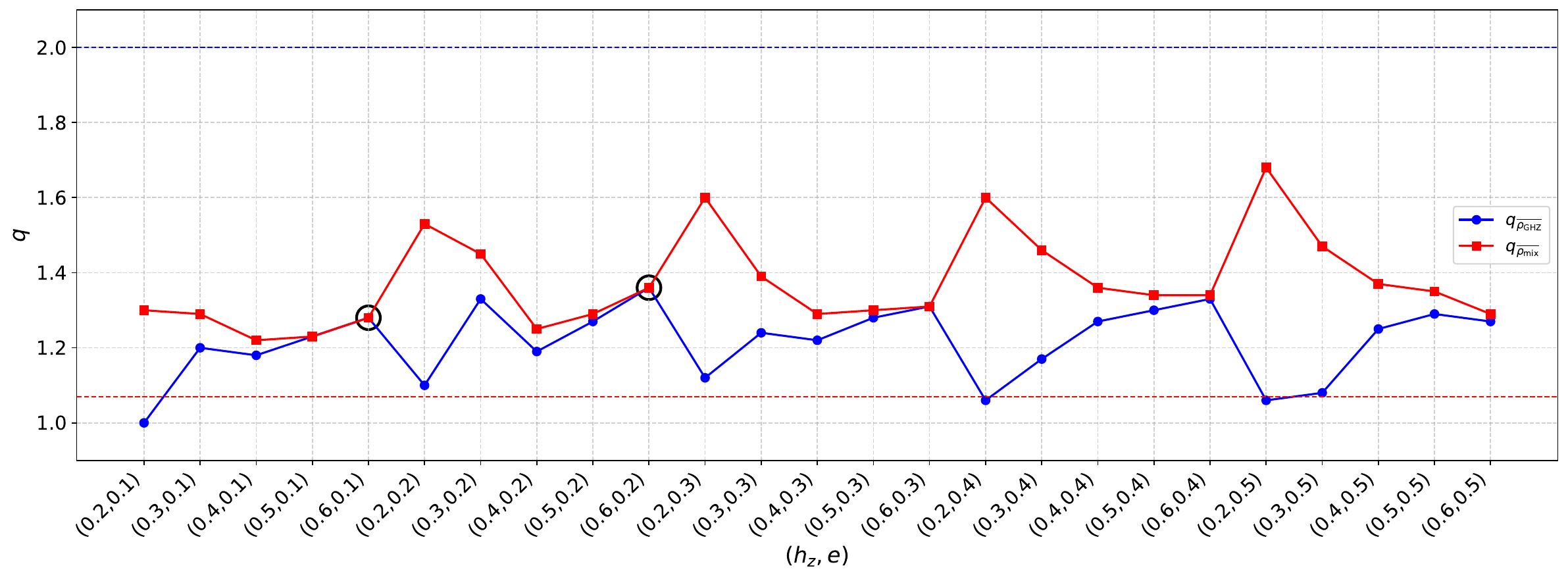}
    \caption{Indices $q$ for different set of parameters, characterized by the choices of $h_{z}$ and $e$. For all the data, $N=10$ (with the exception of the points that are circled, where $N=12$), $q_{\textrm{GHZ}} = 2$ and $q_{\textrm{mix}} = 1.07$ are represented in the dashed lines. Parameters $h_{x}$, $J_{1}$, $J_{2}$, and $d$, are fixed with values 0.2, 1.0, 1.35, and 0.5, respectively.}
    \label{parametros}
\end{figure*}

While Sec.~II showed that equilibration renders a GHZ state and its corresponding classical mixture operationally indistinguishable for a broad class of observables, an important question remains: can the unitary dynamics itself generate new macroscopic coherence in either state? In other words, does equilibration simply hide coherence from accessible measurements, or does it actually suppress macroscopic superpositions altogether? 

To address this, we employ the Shimizu--Miyadera family of measures of macroscopic quantumness and their later extensions ~\cite{RevFrowis,ukena2004,morimae2010,morimae2016, shimizu2018}, which quantify how quantum coherence is distributed across many-body systems. These measures allow us to test whether the long-time states of both the GHZ and the mixture retain any macroscopic coherence. As we show, the infinite-time averages of both states lack macroscopic quantumness, indicating that equilibration does not create new macroscopic superpositions; instead, it progressively erases them.

For a pure state $\rho = \ket{\psi}\bra{\psi}$, macroscopic coherence is quantified by the state index $p$ (with $1 \le p \le 2$) \cite{ukena2004}, defined by how the quantum fluctuations of a normalized local observable scale with system size $N$. The variance of the quantum fluctuation $\Delta A = \text{Tr}[A^2\rho] - \text{Tr}[A\rho]^2 $ maximized over all normalized local observables $A$ satisfies:
\begin{equation}
\underset{\textrm{A:local}}{\textrm{max}} (\Delta A)^2 = O(N^p)  
\label{eq:pure_measure}
\end{equation}
A scaling of $p = 1$ indicates that only short-range correlations are present (each spin correlates with only $O(1)$ neighbors). In contrast, $p > 1$ and, in particular, $p = 2$ signals genuine macroscopic coherence, where quantum correlations extend across an extensive number of pairs.

For mixed states, this measure is insufficient because the variance depends only on diagonal elements of the state in the eigenbasis of $A$. To overcome this, Shimizu and co-authors \cite{shimizu2005} introduced the index $q$, defined by the scaling of the double commutator with a local observable A \cite{morimae2010}:
\begin{equation}
   \textrm{max} \big \{N, \underset{\textrm{A:local}}{\textrm{max}}|| \, [A,[A,\rho]]\, ||_{1} \big \} = O(N^q).  
   \label{eq:mix_measure}
\end{equation}
The "outer" maximization ensures $1 \le q \le 2$. A large trace norm $|| \, [A,[A,\rho]]\, ||_{1} = O(N^2)$ for local operators $A$ is only possible with significant contributions from elements $(a_{k} - a_{l})^2 \langle a_{k}| \rho|a_{l}\rangle = O(N^2)$, which is only possible when coherences extend macroscopically across the system. For pure state, known relations connect the two measures: $q = 1 \Rightarrow p = 1$, $p = 1 \Rightarrow q \le 1.5$ and $q = 2 \Leftrightarrow p = 2$ \cite{morimae2016, shimizu2018}. A state $\rho$ is said to exhibit macroscopic quantum coherence when $q = 2$. States with $q < 2$ may have quantum coherence but not of macroscopic character. States with $q = 1$ resemble separable states from this macroscopic-coherence perspective.

This framework is particularly well suited to our purposes because it characterizes macroscopic distinctness in terms of normalized local additive observables, namely observables of the form $A=\sum_i a_i$ with bounded single-site terms. In this setting, macroscopic coherence is associated with coherence between sectors whose eigenvalues differ by $O(N)$, making this a natural notion for probing genuinely many-body macroscopic superpositions under equilibration. In particular, the magnetization operator introduced previously in Eq.~(\ref{eq:magoperator_main}), is exactly of this form, so the class of observables used here is the same one that already appeared naturally in our operational distinguishability analysis. 

This restriction to additive-local observables is not, by itself, sufficient to guarantee that the measured quantity equilibrates. An additive-local observable may still have conserved quantities, with the Hamiltonian itself being the most direct example. Since such components do not dephase under the unitary dynamics, they can retain persistent contributions for kinematic reasons unrelated to macroscopic coherence. Therefore, in the dynamical use of the indices \(p\) and \(q\) below, the relevant observables should be understood as additive-local probes with no conserved quantities. This is the class of observables for which equilibration is expected and for which the suppression of macroscopic coherence is operationally meaningful.

By contrast, the fully correlated observable $A_{NL}$ discussed in Sec.~II plays a complementary role: it is not part of the additive-local framework underlying the indices $p$ and $q$, but serves to show that even an observable that distinguishes the GHZ state from its corresponding mixture at a fixed time can still lose this distinguishing power under the unitary dynamics. Moreover, in generic many-body systems the detailed unitary dynamics is typically neither fully controlled nor known well enough for an experimenter to tailor measurements to the exact evolution, which further motivates focusing on probes with clear operational meaning under equilibration.

\begin{figure}
    \centering
    \includegraphics[width=\linewidth]{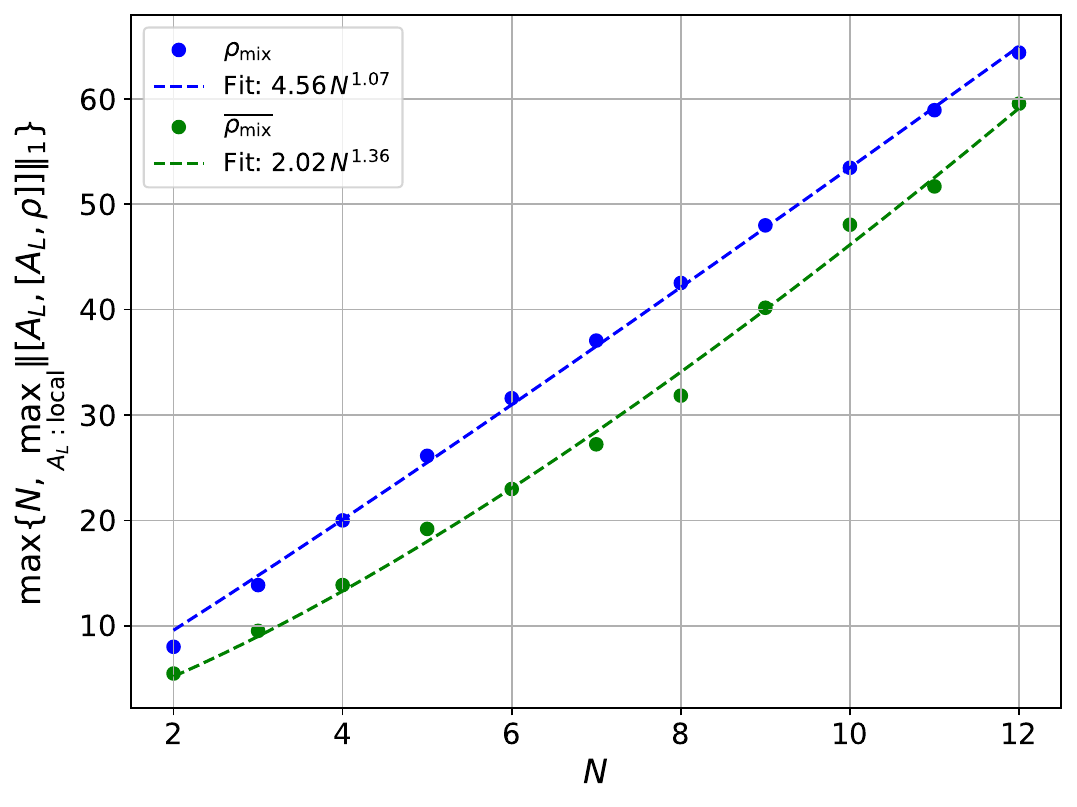} \\[1em]
    \includegraphics[width=\linewidth]{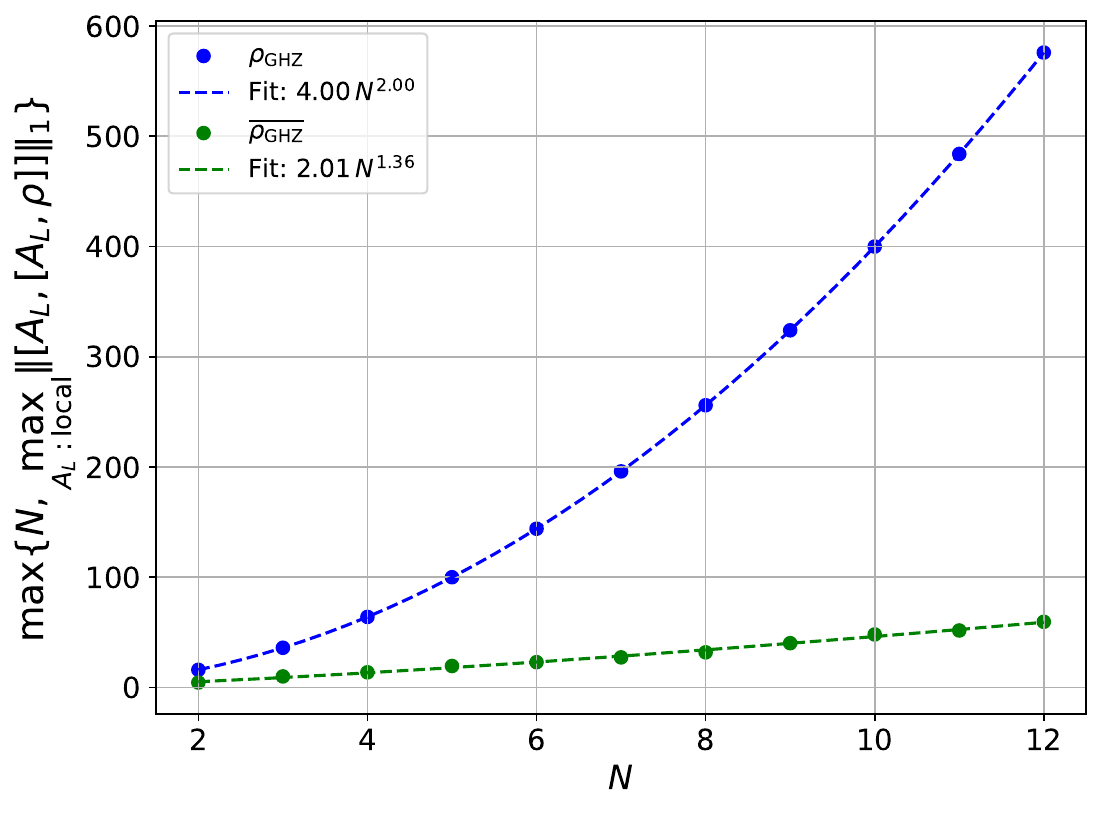} 
    \caption{Obtaining the values of index $q$ for the set of parameters characterized by the second black circle in Figure \ref{parametros}. 
    Top: points representing the measure in Eq. \ref{eq:mix_measure} with $\rho = \rho_{\textrm{mix}}$ and $\rho = \overline{\rho_{\textrm{mix}}}$. 
    Bottom: points representing the in Eq. \ref{eq:mix_measure} with $\rho = \rho_{\textrm{GHZ}}$ and $\rho = \overline{\rho_{\textrm{GHZ}}}$.}
    \label{fig:q_index_second}
\end{figure}

\begin{figure}
    \centering
    \includegraphics[width=\linewidth]{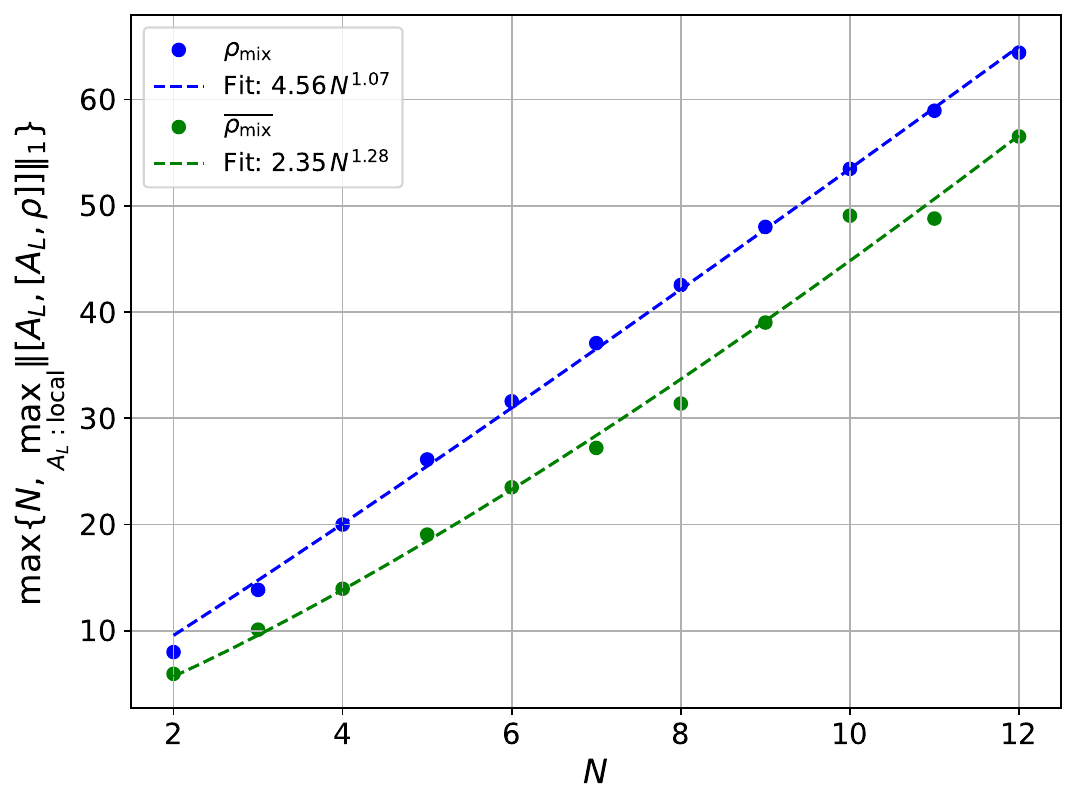} \\[1em]
    \includegraphics[width=\linewidth]{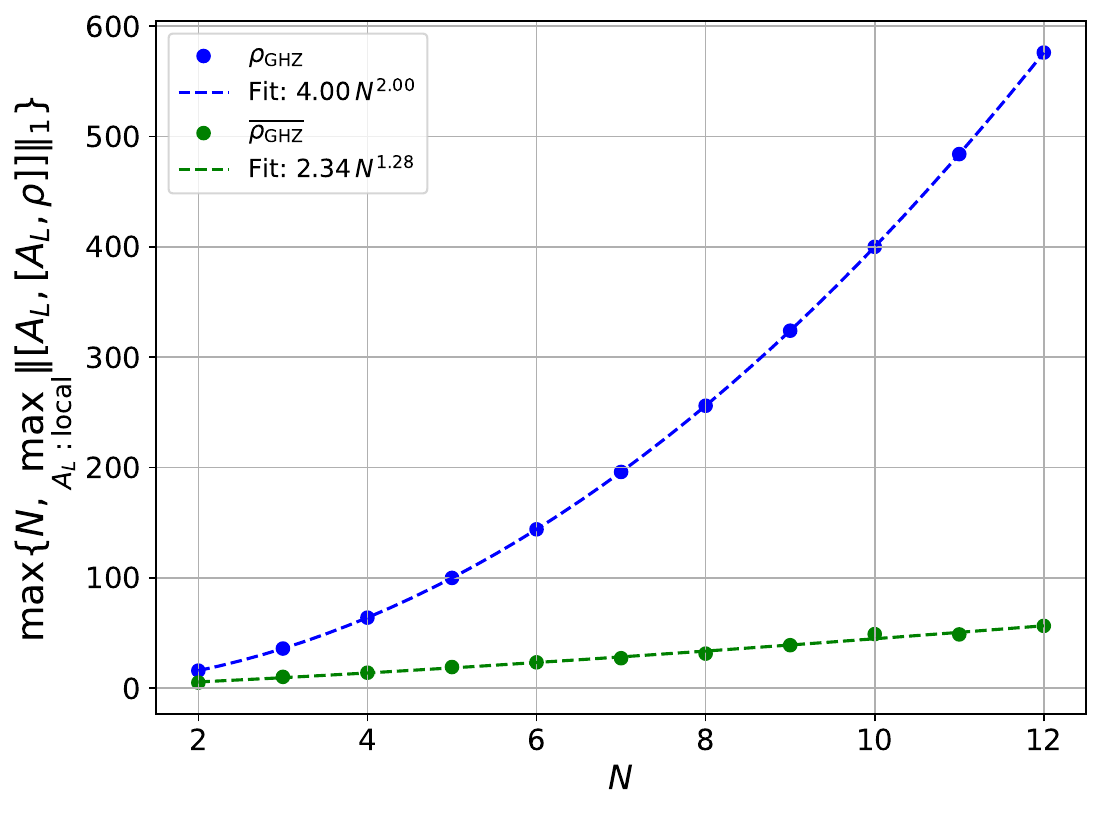} 
    \caption{Obtaining the values of index $q$ for the set of parameters characterized by the first black circle in Figure \ref{parametros}. 
    Top: points representing the measure \ref{eq:mix_measure} with $\rho = \rho_{\textrm{mix}}$ and $\rho = \overline{\rho_{\textrm{mix}}}$. 
    Bottom: points representing the measure \ref{eq:mix_measure} with $\rho = \rho_{\textrm{GHZ}}$ and $\rho = \overline{\rho_{\textrm{GHZ}}}$.}
    \label{fig:q_index_first}
\end{figure}

We evaluated indices $p$ and $q$ for $\rho_{\textrm{GHZ}}$, $\rho_{\textrm{mix}}$, and their time averages $\overline{\rho_{\mathrm{GHZ}}}$ and $\overline{\rho_{\mathrm{mix}}}$ across many parameter sets, maximizing over local operators of the form defined in Eq. (\ref{eq:magoperator_main}). We fixed the parameters $h_{x}$, $J_1$, $J_2$ and $d$, while varying $h_z$ from $0.6$ up to $0.1$ and $e$ from $0.5$ up to $0.1$. For each parameter set, we verified equilibration by confirming a large effective dimension. Figure \ref{parametros} presents the resulting values of $q$ for different choices of $h_z$ and $e$, with systems sizes up to $N=12$. 
Two representative parameter sets, those circled in Figure \ref{parametros} are shown in Figs. \ref{fig:q_index_second} and \ref{fig:q_index_first}, which illustrate how the index $q$ is extracted from Eq. (\ref{eq:mix_measure}). As expected, $p = 2$ for all pure GHZ initial states, so we do not display these values graphically.

After the evolution, the values of $q$ for the macroscopic superposition decrease significantly, whereas the values for the mixture increase, bringing them closer together for all parameter sets. This reflects the fact that the dynamics tends to create moderate coherence in $\rho_{\textrm{mix}}$ and destroy macroscopic coherence in $\rho_{\textrm{GHZ}}$. In both cases, however, the resulting $q$ values remain well below their initial macroscopic value $q = 2$, demonstrating that, within the additive-local framework captured by the index $q$, the long-time states do not retain macroscopic superpositions. In this sense, equilibration suppresses macroscopic quantumness rather than generating it.

\section{Conclusions}
\label{sec:conclusions}

Our analysis shows that although a GHZ state and its corresponding classical mixture differ sharply at $t = 0$--particularly when probed with highly non-local measurements--generic unitary evolution drives them toward operational indistinguishability on experimentally relevant timescales. Even in perfectly isolated systems, the intrinsic
dynamics of a generic many-body Hamiltonian suppresses off-diagonal coherence through dephasing. As a result, both additive-local observables and, in the complementary case analyzed in Sec.~II, the fully correlated observable $A_{NL}$, fail to distinguish macroscopic superpositions from statistical mixtures after equilibration.

More broadly, our findings point to equilibration as a possible ultimate barrier to the observation of macroscopic quantum effects in generic many-body systems. Under mild and physically generic assumptions (nondegenerate gaps, ETH-typical matrix elements, and random-matrix-theory-typical overlaps), the long-time contrast between a macroscopic superposition and its mixture decays as $O(\mathrm{poly}(N)\,2^{-N/2})$, eliminating any operational distinction for experimentally accessible probes. Within the scope considered here, preserving macroscopic coherence or its operational signatures therefore requires violating at least one of the assumptions--for instance, engineering non-generic spectra or constrained dynamics (e.g., many-body localization), employing active quantum error correction, or accessing collective measurements whose effective norm grows super-polynomially. In generic thermalizing systems, any observable difference is confined to finite-size or prethermal regimes, consistent with our numerical simulations for the XYZ model. 

From a physical perspective, it is natural to ask how the conclusions obtained here for an ideal GHZ superposition would change in a more realistic scenario where the macroscopic components are not perfectly pure. In experimental settings, fully polarized configurations such as $\ket{\vec{0}}$ and $\ket{\vec{1}}$ are inevitably replaced by coarse-grained counterparts $\ket{\vec{0}}_{\mathrm{CG}}$ and $\ket{\vec{1}}_{\mathrm{CG}}$, which tolerate a small fraction of spin flips or local defects. These coarse-grained macrostates represent predominantly up- or down-polarized ensembles with polynomially increased impurity and can be viewed as the physical analogues of the constructions introduced in Ref.~\cite{carvalho2024equilibration}. The corresponding superposition, $(\ket{\vec{0}}_{\mathrm{CG}}+\ket{\vec{1}}_{\mathrm{CG}})/\sqrt{2}$, preserves the global structure of a macroscopic quantum state while incorporating realistic microscopic uncertainty. Since the equilibration mechanism depends mainly on the effective dimension and the suppression of off-diagonal coherences, our analysis implies that such coarse-grained GHZ states would also evolve toward operational indistinguishability from their classical mixtures, thereby connecting the macroscopic and coarse-grained descriptions within a unified dynamical picture.

Although our results address typical long-time behavior, several open questions remain. Clarifying the relevant equilibration timescales, identifying symmetry-protected exceptions, and extending this analysis to broader classes of macroscopic or coarse-grained superpositions are natural and compelling next steps. Ultimately, this work establishes equilibration as a fundamental, intrinsic mechanism that enforces a quantum–classical boundary for large, complex systems.

\textit{Acknowledgements.}
This work is supported in part by the National Council for Scientific and Technological Development, 
CNPq Brazil (projects: Universal Grant No. 406499/2021-7, and 409611/2022-0) and it is part of the Brazilian National Institute for Quantum Information. TRO acknowledges funding from the Air Force Office of Scientific Research under Grant No. FA9550-23-1-0092.


%

\end{document}